\documentclass[preprint,times, nofootinbib]{elsarticle}

\usepackage{amsmath}
\usepackage{etoolbox}
\makeatletter
\def\ps@pprintTitle{%
	\let\@oddhead\@empty
	\let\@evenhead\@empty
	\def\@oddfoot{}%
	\let\@evenfoot\@oddfoot}
\makeatother

\newcommand{\ben}[1]{\begin{eqnarray}\label{#1} }
\newcommand{\een}{\end{eqnarray}}

\newcommand{\DD}{{\cal D}}

\newcommand{\eb}{\bar{\epsilon}}

\newcommand{\thc}{\text{h.c}}
\newcommand{\Dslash}{\not{\hbox{\kern-4pt $D$}}}
\newcommand{\pslash}{\not{\hbox{\kern-4pt $\partial$}}}
\newcommand{\Dcslash}{\not{\hbox{\kern-4pt $\DD$}}}

\usepackage{a4wide} 
\usepackage{fancyhdr}
\usepackage{graphicx}
\usepackage{amssymb}
\usepackage{amsthm}
\usepackage{amsmath}
\usepackage{slashed}
\usepackage{amsmath, amsfonts, amssymb, array}
\usepackage{mathrsfs}
\usepackage{slashed}
\newcolumntype{C}[1]{>{\centering\let\newline\\\arraybackslash\hspace{0pt}}m{#1}}
\usepackage{color}
\usepackage[latin1]{inputenc} 
\usepackage[T1]{fontenc} 
\usepackage{multirow}
\usepackage[english]{babel}
\usepackage{bbold}
\usepackage{blkarray}
\usepackage{multirow}

\usepackage[hidelinks]{hyperref}

\hypersetup{
	colorlinks   = true, 
	urlcolor     = blue, 
	linkcolor    = blue, 
	citecolor   = blue 
}

\journal{}

\begin{document}
	
	\begin{frontmatter}
		
		\title{Comment on "The $\mathcal{N} = 3$ Weyl Multiplet in Four Dimensions"}
		\author{Subramanya Hegde\footnote{Corresponding author. Email: smhegde14@iisertvm.ac.in} and Bindusar Sahoo}
		\address{			Indian Institute of Science Education and Research Thiruvananthapuram, Vithura, Kerala, 695551, India}

		\begin{abstract}
		$\mathcal{N}=3$ Weyl multiplet in four dimensions was first constructed in J van Muiden et al (2017) where the authors used the current multiplet approach to obtain the linearized transformation rules and completed the non-linear variations using the superconformal algebra. The multiplet of currents was obtained by a truncation of the multiplet of currents for the $\mathcal{N}=4$ vector multiplet. While the procedure seems to be correct, the result suffers from several inconsistencies. The inconsistencies are observed in the transformation rules as well as the field dependent structure constants in the corresponding soft algebra. We take a different approach, and compute the transformation rule as well as the corresponding soft algebra by demanding consistency.
		\end{abstract}
		
	\end{frontmatter}

\allowdisplaybreaks
\baselineskip 3.5ex
\section{Introduction}
$\mathcal{N}$ extended conformal supergravity in four dimensions is a theory of gravity where the fields form a representation of the $su(2,2|\mathcal{N})$ superconformal algebra. While the resulting theory is not physical, the additional conformal symmetries can be gauge fixed by use of compensating multiplets to obtain matter coupled Poincar\'e supergravity theories. Construction of multiplets in conformal supergravity as well as their action, is facilitated by a set of procedures which are collectively termed as superconformal multiplet calculus. It provides us a systematic method to construct matter couplings in Poincar\'e supergravity. In particular, the use of off-shell multiplets enables the construction of general matter couplings, which has been utilized to construct quadratic as well as higher derivative invariants in Poincar\'e supergravity, see e.g.
\cite{Bergshoeff:1980is,deWit:2010za,Butter:2013lta,Kuzenko:2015jxa,Hanaki:2006pj,Ozkan:2013nwa,Butter:2014xxa} .

A multiplet in conformal supergravity which contains all the gauge fields of the superconformal algebra is known as the Weyl multiplet. It has been found that there can be more than one such Weyl multiplet, in any given dimension ($d\le6$), with different auxiliary field content \cite{Bergshoeff:1985mz,Bergshoeff:2001hc,Muller:1986ts,Muller:1986ku,Siegel:1995px,Butter:2017pbp}. The Weyl multiplet which consists of a real scalar field of Weyl weight $+1$ is said to be the dilaton Weyl multiplet, and the other Weyl multiplet is known as the standard Weyl multiplet. To construct these multiplets, one often takes recourse to the current multiplet method. The multiplet of currents for a rigid on-shell multiplet is computed and it is then coupled to fields via a first order action to obtain the linearized transformation rules. Non-linear transformation rules are then obtained by adding all possible terms to the transformation rule consistent with their Weyl weight, chiral weight, Lorentz and $R$-symmetry structure and demanding consistency with the superconformal algebra. In this process, to realize this as a theory of gravity where local translations act as general coordinate transformations, one has to impose constrains on some of the conformal curvatures which renders the algebra to be a soft algebra i.e., the algebra is satisfied upto field dependent gauge transformations. In four dimensions, this process has been used to derive the transformation rule for the $\mathcal{N}=2$\cite{deWit:1979dzm} and $\mathcal{N}=4$\cite{Bergshoeff:1980is} Weyl multiplets as well as to determine their soft algebra. 

However, in the case of $\mathcal{N}=3$ supersymmetry, it is known that any multiplet with particles of spin less than $2$, also admits a fourth supersymmetry. Thus if we begin with a two derivative action for a rigid $\mathcal{N}=3$ vector multiplet, the resulting current multiplet will be equivalent to the $\mathcal{N}=4$ current multiplet upto field redefinitions. In \cite{vanMuiden:2017qsh}, this obstruction was overcome by performing a truncation of the $\mathcal{N}=4$ current multiplet to the $\mathcal{N}=3$ case. The truncated $\mathcal{N}=3$ current multiplet then gave the linearized transformations for the $\mathcal{N}=3$ Weyl multiplet as well as the soft algebra on imposing the conventional constraints.\footnote{For earlier works on $\mathcal{N}=3$ supergravity using other approaches, see \cite{Rivelles:1981qy,Brink:1978iv,Galperin:1986id,Castellani:1985ka}} 

However, the results of \cite{vanMuiden:2017qsh} has several inconsistencies both at the linearized as well as the nonlinear level for the transformation rule as well as the field dependent gauge transformation parameters of the soft algebra.  

In this letter, we compute the transformation rule of the $\mathcal{N}=3$ Weyl multiplet as well as the field dependent gauge transformation parameters appearing in the soft algebra.

\section{Corrected transformation rules and the algebra}
We will first demonstrate, by an example, the inconsistencies found in the results of \cite{vanMuiden:2017qsh}. The field content of the $\mathcal{N}=3$ Weyl multiplet is given in Table-\ref{Table-Weyl}, with appropriate Weyl weights, chiral weights\footnote{Field content for the $\mathcal{N}=3$ Weyl multiplet were first suggested in \cite{Fradkin:1985am,vanNieuwenhuizen:1985dp}}. Let us consider linear terms in the transformation of the fields $T_{abj}, V_\mu{}_j{}^i, A_\mu$, which contain $\zeta^i$. To compute this we will use the following transformations, where we have fixed the coefficients by chosing a normalization for the fields $e_\mu^a, \psi_\mu^i, T_{ab j}$ and $\Lambda_L$.
\begin{table}[t]
	\caption{Field content of the $N=3$ Weyl multiplet}\label{Table-Weyl}
	\begin{center}
		\begin{tabular}{ | C{2cm}|C{2cm}|C{3cm}|C{2cm}|C{2cm}| }
			\hline
			Field & SU(3) Irreps & Restrictions &Weyl weight (w) & Chiral weight (c) \\ \hline
			$e_{\mu}{}^{a}$ & $\bf{1}$ & Vielbein & -1 & 0 \\ \hline
			$V_{\mu}{}^{i}{}_{j}$ & $\bf{8}$ & $(V_{\mu}{}^{i}{}_{j})^{*}\equiv V_{\mu}{}_{i}{}^{j}=-V_{\mu}{}^{j}{}_{i}$ SU(4)$_R$ gauge field &0 & 0  \\ \hline
			$A_{\mu}$ & $\bf{1}$ & U(1)$_R$ gauge field &0 & 0  \\ \hline
			$b_{\mu}$ & $\bf{1}$ & dilatation gauge field &0 & 0  \\ \hline
			$T^{i}_{ab}$ & $\bf{3}$ & Self-dual i.e $T^{i}_{ab}=\frac{1}{2}\varepsilon_{abcd}T^{i}{}^{cd}$ &1 & -1  \\ \hline
			$E_{i}$ &$\bf{\bar{3}}$ & Complex & 1&1\\ \hline
			$D^{i}{}_{j}$ & $\bf{8}$ & $(D^{i}{}_{j})^{*}\equiv D_{i}{}^{j}=D^{j}{}_{i}$ &2 & 0  \\ \hline
			$\psi_{\mu}{}^{i}$ & $\bf{3}$ & $\gamma_{5}\psi_{\mu}{}^{i}=\psi_{\mu}{}^{i}$&-1/2 & +1/2  \\ \hline
			$\chi_{ij}$ & $\bf{\bar{6}}$ & $\gamma_{5}\chi_{ij}=\chi_{ij}$ &3/2 &+1/2  \\ \hline
			$\zeta^{i}$ & $\bf{3}$ & $\gamma_{5}\zeta^{i}=\zeta^{i}$ & 3/2 &+1/2 \\ \hline
			$\Lambda_{L}$ & $\bf{1}$ & $\gamma_{5}\Lambda_{L}=\Lambda_{L}$ &1/2 &+3/2 \\ \hline
		\end{tabular}
	\end{center}
\end{table}
\begin{align}
\delta e_{\mu}^{a}&=\frac{1}{2}\bar{\epsilon}_{i}\gamma^{a}\psi_{\mu}^{i}+\thc\nonumber\\
\delta \psi_{\mu}^{i}&=\mathcal{D}_{\mu}\epsilon^{i}-\frac{1}{4}\varepsilon^{ijk}\gamma\cdot T_{j}\gamma_{\mu}\epsilon_{k}-\varepsilon^{ijk}\bar{\epsilon}_{j}\psi_{\mu k}\Lambda_{L}-\gamma_{\mu}\eta^{i}
\end{align}
Let us write the possible terms in the variation of $T_{abj}, V_\mu{}_j{}^i, A_\mu$ linear in $\zeta_i$ with arbitrary coefficients.
\begin{align}
\delta V_\mu{}_j{}^i &=\alpha_1\bar{\epsilon}^i\gamma_\mu\zeta_j-\thc-\text{trace}\nonumber\\
\delta A_\mu &= i\alpha_2\bar{\epsilon}^i\gamma_\mu\zeta_i+\thc\nonumber\\
\delta T_{ab}^i &=\alpha_3\varepsilon^{ijk}\bar{\epsilon}_j\gamma_{ab}\zeta_k
\end{align}
The coefficients $\alpha_1, \alpha_2$ and $\alpha_3$ can be determined from the $Q-Q$ algebra on $\psi_\mu^i$. Accordingly, we obtain the relations $\alpha_2=-4 \frac{\alpha_1}{3}$ and $\alpha_3=\frac{\alpha_1}{2}$ while $\alpha_1$ can be fixed for a normalization of $\zeta^i$. On the R.H.S. of the algebra we obtain,
\begin{align}
[\delta^{Q}(\epsilon_1),\delta^Q (\epsilon_2)]\psi_\mu^i&=-\alpha_1(\bar{\epsilon}_1^i\epsilon_2^j-\bar{\epsilon}_2^i\epsilon_1^j)\gamma_\mu\zeta_j\nonumber\\
&\quad+\frac{\alpha_1}{4}\gamma_\mu(\bar{\epsilon}_1^i\gamma^{a}\epsilon_{2j}-\delta^i_j\bar{\epsilon}_1^k\gamma^{a}\epsilon_{2k}+\thc)\gamma_a\zeta^j
\end{align}
Which is a field dependent $S$-transformation $\delta_S(\eta_a^j)\psi_\mu^i=-\gamma_{\mu}\eta^i$ with the parameter,
\begin{align}
\eta_1^i&=\alpha_1(\bar{\epsilon}_1^i\epsilon_2^j-\bar{\epsilon}_2^i\epsilon_1^j)\zeta_j\nonumber\\
&\quad-\frac{\alpha_1}{4}(\bar{\epsilon}_1^i\gamma^{a}\epsilon_{2j}-\delta^i_j\bar{\epsilon}_1^k\gamma^{a}\epsilon_{2k}+\thc)\gamma_a\zeta^j
\end{align}
However, the results in \cite{vanMuiden:2017qsh} read $\delta V_\mu{}_j{}^i=\frac{1}{12}\bar{\epsilon}^i\gamma_\mu\zeta_j-\thc-\text{trace}$, $\delta A_\mu=-\frac{i}{9}\bar{\epsilon}^i\gamma_\mu\zeta_i+\thc$, $\delta T_{ab}^i=\frac{1}{12}\varepsilon^{ijk}\bar{\epsilon}_j\gamma_{ab}\zeta_k$ and $\eta_{1}^i=\frac{3}{4}(\bar{\epsilon}_1^i\epsilon_2^j-\bar{\epsilon}_2^i\epsilon_1^j)\zeta_j + \frac{3}{2}(\bar{\epsilon}_1^i\gamma^{a}\epsilon_{2j}-\delta^i_j\bar{\epsilon}_1^k\gamma^{a}\epsilon_{2k}+\thc)\gamma_a\zeta^j$ which is clearly not consistent with the coefficients we obtain from the algebra. Thus, the inconsistencies already exist at the linearized level in the transformation rule as well as the field dependent gauge transformation parameters appearing in the soft algebra. As numerous such inconsistencies exist in the result, it calls for a reconstruction of the multiplet to render it useful for future purposes.

We have performed such a reconstruction, as demonstrated in the example above, solely by demanding consistency with the superconformal algebra while allowing field dependent gauge transformations to appear on the right hand side of the soft algebra.

We have given the $Q$ and $S$ transformation of the fields below.
\begin{align}
\delta e_{\mu}^{a}&=\frac{1}{2}\bar{\epsilon}_{i}\gamma^{a}\psi_{\mu}^{i}+\thc\nonumber\\
 \delta \psi_{\mu}^{i}&=\mathcal{D}_{\mu}\epsilon^{i}-\frac{1}{4}\varepsilon^{ijk}\gamma\cdot T_{j}\gamma_{\mu}\epsilon_{k}-\varepsilon^{ijk}\bar{\epsilon}_{j}\psi_{\mu k}\Lambda_{L}-\gamma_{\mu}\eta^{i}\nonumber\\
 \delta V_\mu{}_j{}^i &=-\bar{\epsilon}^i\phi_{\mu j}+\frac{1}{12}\bar{\epsilon}^i\gamma_\mu\zeta_j-\frac{1}{4}\varepsilon_{jkl}\bar{\epsilon}^k\gamma_\mu\chi^{il}+\frac{1}{4}\bar{\epsilon}^i\gamma\cdot T_j \gamma_\mu\Lambda_R+\frac{1}{4}\bar{\epsilon}^i\gamma_\mu \Lambda_R E_j -\frac{1}{4}\varepsilon_{klj}E^i\bar{\epsilon}^k\psi_\mu^l\nonumber\\
&\quad-\frac{1}{2}\bar{\epsilon}^i\gamma^a\psi_{\mu j}\bar{\Lambda}_L\gamma_a\Lambda_R+\bar{\psi}_\mu^i\eta_j-\thc-\text{trace}\nonumber\\
\delta A_\mu &=-\frac{i}{6}\bar{\epsilon}^i\phi_{\mu i}-\frac{i}{9}\bar{\epsilon}^i\gamma_\mu\zeta_i-\frac{i}{6}\varepsilon_{klp}E^p\bar{\epsilon}^k\psi_{\mu}^l-\frac{i}{3}\bar{\epsilon}^i\gamma\cdot T_i\gamma_\mu\Lambda_R-\frac{i}{3}\bar{\epsilon}^i\gamma_\mu\Lambda_RE_i+\frac{2i}{3}\bar{\epsilon}^i\gamma^a\psi_{\mu i}\bar{\Lambda}_L\gamma_a\Lambda_R\nonumber\\
&\quad+\frac{i}{6}\bar{\psi}_\mu^i\eta_i+\thc\nonumber\\
\delta b_\mu &= \frac{1}{2}(\bar{\epsilon}^i\phi_{\mu i}-\bar{\psi}_\mu^i\eta_i)+\thc\nonumber\\
\delta \Lambda_L&=-\frac{1}{4}E_i\epsilon^i+\frac{1}{4}\gamma\cdot T_i\epsilon^i\nonumber\\
\delta E_i &=-\bar{\epsilon}_i\slashed{D}\Lambda_L-\frac{1}{2}\varepsilon_{ijk}\bar{\epsilon}^j\zeta^k+\frac{1}{2}\bar{\epsilon}^j\chi_{ij}-\frac{1}{2}\varepsilon_{ijk}E^k\bar{\epsilon}^j\Lambda_L-4\bar{\Lambda}_L\Lambda_L\bar{\epsilon}_i\Lambda_R-2\bar{\eta}_i\Lambda_L\nonumber\\
\delta T^i_{ab} &= -\frac{1}{4}\bar{\epsilon}^i\slashed{D}\gamma_{ab}\Lambda_R-\frac{1}{2}\varepsilon^{ijk}\bar{\epsilon}_jR_{ab}(Q)_k+\frac{1}{8}\bar{\epsilon}_j\gamma_{ab}\chi^{ij}+\frac{1}{24}\varepsilon^{ijk}\bar{\epsilon}_j\gamma_{ab}\zeta_k-\frac{1}{8}\varepsilon^{ijk}E_j\bar{\epsilon}_k\gamma_{ab}\Lambda_R\nonumber\\
&\quad+\frac{1}{2}\bar{\eta}^i\gamma_{ab}\Lambda_R\nonumber\\
\delta \chi_{ij}&=\frac{1}{2}\slashed{D}E_{(i}\epsilon_{j)}+\frac{1}{2}\varepsilon_{kl(i}\gamma\cdot R(V)_{j)}{}^l\epsilon^k-\frac{1}{2}\gamma\cdot\slashed{D}T_{(i}\epsilon_{j)}+\frac{1}{3}\varepsilon_{kl(i}D^l{}_{j)}\epsilon^k\nonumber \\&\quad+\frac{1}{4}\varepsilon_{kl(i}E^k\gamma\cdot T_{j)}\epsilon^l-\frac{1}{3}\bar{\Lambda
}_L\gamma_a\epsilon_{(i}\gamma^a\zeta_{j)}+\frac{1}{4}\varepsilon_{lm(i}E_{j)}E^m\epsilon^l-\bar{\Lambda}_L\gamma^a\Lambda_R\gamma_aE_{(i}\epsilon_{j)}\nonumber\\
&\quad-\bar{\Lambda}_L\gamma\cdot T_{(i}\gamma^a\Lambda_R\gamma_a\epsilon_{j)}+\gamma\cdot T_{(i}\eta_{j)}+E_{(i}\eta_{j)}\nonumber\\
\delta \zeta^i &=-\frac{3}{4}\varepsilon^{ijk}\slashed{D}E_j\epsilon_k+\frac{1}{4}\varepsilon^{ijk}\gamma\cdot\slashed{D}T_k\epsilon_j+\frac{1}{4}\gamma\cdot R(V)_j{}^i\epsilon^j+i\gamma\cdot R(A)\epsilon^i-\frac{1}{2}D^i{}_j\epsilon^j-\frac{3}{8}E^i\gamma\cdot T_j\epsilon^j\nonumber\\
&\quad+\frac{3}{8}E^j\gamma\cdot T_j\epsilon^i+\frac{3}{8}E^iE_j\epsilon^j+\frac{1}{8}E^jE_j\epsilon^i\nonumber\\
&\quad-\bar{\Lambda}_L\slashed{D}\Lambda_{R}\epsilon^i-\bar{\Lambda}_R\slashed{D}\Lambda_L\epsilon^i-\frac{3}{4}\bar{\Lambda}_R\slashed{D}\gamma_{ab}\Lambda_L\gamma^{ab}\epsilon^i-\frac{3}{4}\bar{\Lambda}_L\gamma_{ab}\slashed{D}\Lambda_R\gamma^{ab}\epsilon^i\nonumber\\
&\quad+\frac{1}{2}\varepsilon^{ijk}\bar{\Lambda}_L\gamma^a\epsilon_j\gamma_a\zeta_k-6\bar{\Lambda}_L\Lambda_L\bar{\Lambda}_R\Lambda_R\epsilon^i+\frac{1}{2}\varepsilon^{ijk}\gamma\cdot T_j\eta_k-\frac{3}{2}\varepsilon^{ijk}E_j\eta_k\nonumber\\
\delta D^i_j&=-\frac{3}{4}\bar{\epsilon}^i\slashed{D}\zeta_j-\frac{3}{4}\varepsilon_{jkl}\bar{\epsilon}^k\slashed{D}\chi^{il}+\frac{1}{4}\varepsilon_{jkl}\bar{\epsilon}^i\zeta^k E^l+\frac{1}{2}\varepsilon_{jkl}\bar{\epsilon}^k\zeta^l E^i+\frac{3}{4}\bar{\epsilon}^i\chi_{jk}E^k+\frac{3}{4}\bar{\epsilon}^i\gamma\cdot T_j\overset{\leftrightarrow}{\slashed{D}}\Lambda_R\nonumber\\
&\quad-\frac{3}{4}\bar{\epsilon}^i\slashed{D}\Lambda_RE_j-\frac{3}{4}\bar{\epsilon}^i\slashed{D}E_j\Lambda_R+\frac{3}{4}\varepsilon_{jkl}E^l\bar{\epsilon}^k\Lambda_LE^i+3\varepsilon_{jkl}T^i\cdot T^l\bar{\epsilon}^k\Lambda_{L}-2\bar{\epsilon}^i\Lambda_L\bar{\Lambda}_R\zeta_j-3\bar{\epsilon}^i\Lambda_L\bar{\Lambda}_R\Lambda_RE_j\nonumber\\
&\quad+3\bar{\epsilon}^i\gamma\cdot T_j\Lambda_L\bar{\Lambda}_R\Lambda_R+\thc-\text{trace}
\end{align}
The supersymmetry algebra takes the following form:
 \begin{align}
[\delta^{Q}(\epsilon_1),\delta^Q (\epsilon_2)]&=\delta^{cgct}(\xi^{\mu})+\delta^{M}(\epsilon_{1}^{ab})+\delta^{Q}(\epsilon_{3}^{i})+\delta^{S}(\eta_{1}^{i})+\delta_{SU(3)}(\lambda_{1}{}_{j}{}^{i})+\delta_{U(1)}(\lambda_{1T})+\delta_K(\lambda^a_{1K})\nonumber \\
[\delta^{Q}(\epsilon),\delta^S (\eta)]&= \delta_{D}(\lambda_D)+\delta^{M}(\epsilon_{2}^{ab})+\delta^{S}(\eta_{2}^{i})+\delta_{SU(3)}(\lambda_{2}{}_{j}{}^{i})+\delta_{U(1)}(\lambda_{2T})+\delta_K(\lambda^a_{2K})\nonumber\\
[\delta^{S}(\eta_1),\delta^S (\eta_2)]&=\delta_K(\bar{\eta}_2^i\gamma^a\eta_{1i}+\thc)
 \end{align}
 The parameters appearing above are:
 \begin{align}
 \xi^{\mu}&=\frac{1}{2}\eb_{2i}\gamma^{\mu}\epsilon_{1}^{i}+\thc\nonumber \\
 \epsilon_{1}^{ab}&=-\varepsilon_{ijk}\eb_{2}^{i}\epsilon_{1}^{j}T_{ab}^{k}+\thc\nonumber \\
 \epsilon_{3}^{i}&=-\varepsilon^{ijk}\eb_{2j}\epsilon_{1k}\Lambda_L\nonumber \\
 \eta_{1}^{i}&=-\frac{1}{6}\eb_2^{[i}\epsilon_1^{k]}\zeta_{k}+\frac{1}{16}\left(\eb_{2}^{i}\gamma_{a}\epsilon_{1j}-\delta^{i}_{j}\eb_{2}^{k}\gamma_{a}\epsilon_{1k}+h,c\right)\gamma^{a}\Lambda_{L}E^{j}\nonumber\\
 &\quad+\frac{1}{48}\left(\eb_{2}^{i}\gamma_{a}\epsilon_{1j}-\delta^{i}_{j}\eb_{2}^{k}\gamma_{a}\epsilon_{1k}+h,c\right)\gamma^{a}\zeta^j-\frac{1}{2}\bar{\epsilon}_2^{[i}\epsilon_1^{j]}E_j\Lambda_R\nonumber\\
 &\quad-\frac{1}{16}\varepsilon^{ijk}(\bar{\epsilon}_2^l\gamma_a\epsilon_{1k}+\thc)\gamma^a\chi_{jl}-\frac{1}{4}\varepsilon_{jkl}\bar{\epsilon}_2^j\epsilon_1^k\chi^{il}-\frac{1}{4}\varepsilon^{ijk}\bar{\epsilon}_{2j}\epsilon_{1k}\slashed{D}\Lambda_L\nonumber\\
 &\quad-\frac{1}{8}(\eb_2^i\gamma_a\epsilon_{1j}-\delta^i_j \eb_2^k\gamma_a\epsilon_{1k}+\thc)\gamma\cdot T^j\gamma^a\Lambda_L\nonumber\\ 
 \lambda_{1T}&=-\frac{i}{6}\varepsilon_{ijk}\bar{\epsilon}_2^j\epsilon_{1}^kE^i+\frac{2i}{3}(\bar{\epsilon}_2^j\gamma^a\epsilon_{1j})\bar{\Lambda}_L\gamma_a\Lambda_R+\thc\nonumber\\
 \lambda_1{}_j{}^i &=-\frac{1}{4}\varepsilon_{jpq}\bar{\epsilon}_2^p\epsilon_1^q E^i-\frac{1}{2}(\bar{\epsilon}_2^i\gamma^a\epsilon_{1j})\bar{\Lambda}_L\gamma_a\Lambda_R-\thc-\text{trace}\nonumber\\
  \lambda^a_{1K}&=\frac{i}{12}(\eb_{2}^{k}\gamma_{b}\epsilon_{1k}-\eb_{1}^{k}\gamma_{b}\epsilon_{2k})\tilde{R}^{ab}(A)-\frac{1}{6}(\eb_{2}^{i}\gamma_{b}\epsilon_{1j}-\eb_{1}^{i}\gamma_{b}\epsilon_{2j})\tilde{R}^{ab}(V){}_i{}^j\nonumber\\
 &\quad+\frac{1}{8}(\eb_2^{[i}\gamma\cdot T_i\gamma_{\mu}\gamma\cdot T^{j]}\epsilon_{1j}-\eb_1^{[i}\gamma\cdot T_i\gamma_{\mu}\gamma\cdot T^{j]}\epsilon_{2j})-\frac{2}{3}\varepsilon^{ijk}\eb_{2i}\epsilon_{1j}D_b T^{ab}_k\nonumber\\
 \lambda_D &= -\frac{1}{2}\bar{\eta}_i\epsilon^i+\thc \nonumber \\
\epsilon_2^{ab} &=\frac{1}{2}\bar{\eta}_i\gamma^{ab}\epsilon^i+\thc\nonumber\\
\eta_{2i}&=\frac{1}{4}\varepsilon_{ijk}\bar{\epsilon}^j\gamma_a\eta^k\gamma^a\Lambda_R\nonumber\\
\lambda_2{}_j{}^i &=\bar{\epsilon}^i\eta_j-\thc-\text{trace}\nonumber\\
\lambda_{2T} &=\frac{i}{6}\bar{\epsilon}^i\eta_i+\thc\nonumber\\
\lambda_{2K}^a&=-\frac{1}{12}\varepsilon_{ijk}\eb^i\gamma^a\gamma\cdot T^j\eta^k+\thc
 \end{align}
Here note that, among the independent fields, only $b_a$ transforms under $K$ transformations as $\delta_K b^a=\lambda_K^a$. Covriant derivative on the supersymmetry parameter $\epsilon^i$ is defined as,
\begin{align}
D_\mu\epsilon^i=\partial_\mu\epsilon^i+\frac{1}{4}\gamma\cdot\omega_\mu\epsilon^i+\frac{1}{2}(b_\mu-iA_\mu)\epsilon^i+V_\mu{}_j{}^i\epsilon^j 
\end{align}

Transformation of the dependent gauge fields $\omega_\mu^{ab}$ and $\phi_\mu^i$ is given as follows.
\begin{align}
\delta \omega_\mu^{ab}&=\frac{1}{2}\bar{\epsilon}^i\gamma^{ab}\phi_{\mu i}-\varepsilon_{ijk}\bar{\epsilon}^i\psi_\mu^jT^{ab k}-\frac{1}{2}\bar{\epsilon}^i\gamma_\mu R(Q)^{ab}{}_i+\frac{1}{2}\bar{\eta}^i\gamma^{ab}\psi_{\mu i}+\thc\nonumber\\
\delta_{\text{cov}}\phi_\mu^i&=\frac{i}{48}(\gamma_\mu\gamma\cdot R(A)-3\gamma\cdot R(A)\gamma_\mu)\epsilon^i+\frac{1}{24}(3\gamma\cdot R(V){}_j{}^i\gamma_\mu-\gamma_\mu\gamma\cdot R(V){}_j{}^i)\epsilon^j\nonumber\\
&\quad-\frac{1}{8}\varepsilon^{ijk}\bar{\Lambda}_L\gamma_{\mu}R_{ab}(Q)_k\gamma^{ab}\epsilon_j+\frac{1}{8}\gamma\cdot T^{[i}\gamma_{\mu}\gamma\cdot T_j\epsilon^{j]}+\frac{1}{24}\varepsilon^{ijk}(\gamma_{\mu}\gamma\cdot\slashed{D}T_j-3\slashed{D}\gamma\cdot T_j\gamma_\mu)\epsilon_k\nonumber\\
&\quad-\frac{1}{12}\varepsilon^{ijk}\gamma_\mu\gamma\cdot T_j\eta_k
\end{align}
where we have only given the covariant terms in the transformation of $\phi_\mu^i$. Transformation of the rest of the dependent gauge fields, as well as the curvatures can be found by using the techniques given in \cite{Freedman:2012zz}.

We can see from the above that several terms appear here with non zero coefficients which were not present in the transformation rule of \cite{vanMuiden:2017qsh} while other terms are corrected to different coefficients. The field dependent gauge transformation parameters appearing in the soft algebra are also correted. 

\section{Conclusions}
In this letter, we have presented the transformation rule for the $\mathcal{N}=3$ Weyl multiplet, along with the field dependent gauge transformation parameters that appear in the soft algebra. We have indicated, by an example, the inconsistencies that exist in the results of \cite{vanMuiden:2017qsh}. We would like to emphasize that such errors propagate by the nature of the current multiplet construction where the linear terms are used to fix the coefficients of the nonlinear terms. Hence, we hope our reconstruction will be of value for further use of this multiplet and hence is of interest to the community.

The result opens up the possibility of finding matter coupled $\mathcal{N}=3$ Poincar\'e supergravity theories, which has not been done so far using the superconformal method. In particular, one can use the techniques of \cite{Butter:2016mtk} to construct the fully nonlinear action for $\mathcal{N}=3$ conformal supergravity, which is a work in progress. Such a construction would then allow construction of higher derivative invariants in $\mathcal{N}=3$ Poincar\'e supergravity. Another possibility is also the construction of the dilaton Weyl multiplet in $\mathcal{N}=3$ conformal supergravity by coupling the above multiplet to an on-shell $\mathcal{N}=3$ vector multiplet and construction of the Poincar\'e supergravity invariants by use of the dilaton Weyl multiplet.

\section*{Acknowledgements}
We thank Antoine Van Proeyen and Jesse van Muiden for a discussion on the results and useful comments on the manuscript.

We thank Amitabh Virmani for hospitality at the Chennai Mathematical Institute where a part of this work was completed. We express our gratitude to people of India for their continued support to the basic sciences.

\bibliographystyle{elsarticle-num}
\bibliography{references}
\end{document}